\documentclass[todotoc]{PoS}
\usepackage{amssymb}
\usepackage{amsmath}

\title{Analogue spacetimes: \\ Toy models for ``quantum gravity''}

\ShortTitle{Analogue spacetimes: Toy models for ``quantum gravity''}

\author{\speaker{Matt Visser}\thanks{Supported by the Marsden Fund administered by the Royal Society of New Zealand.} \,
and Silke Weinfurtner\thanks{Current address:  Dept of Physics and Astronomy, University of British Columbia, Vancouver, BC~V6T~1Z1, Canada. Partially supported by a VUW PhD completion scholarship, partially supported by the Marsden Fund.}%
         \\
        School of Mathematics, Statistics, and Computer Science,\\
Victoria University of Wellington, New Zealand\\
        E-mail: \email{\{matt.visser,silke.weinfurtner\}@mcs.vuw.ac.nz}}

\abstract{
Why are ``analogue spacetimes'' interesting? For the purposes of this workshop the answer is simple: Analogue spacetimes provide one with physically well-defined and physically well-understood concrete models of many of the phenomena that seem to be part of the yet incomplete theory of ``quantum gravity'', or more accessibly, ``quantum gravity phenomenology''.  
Indeed  ``analogue spacetimes'' provide one with concrete models of   ``emergence''  (whereby the effective low-energy theory can be \emph{radically} different from the high-energy  microphysics). They also provide many concrete and controlled models of ``Lorentz symmetry breaking'', and extensions of the usual notions of pseudo-Riemannian geometry such as ``rainbow spacetimes'', and pseudo--Finsler geometries, and more. I will provide an overview of the key items of ``unusual physics''  that arise in analogue spacetimes, and argue that they provide us with \emph{hints} of what we should be looking for in any putative theory of ``quantum gravity''.  
For example: The dispersion relations that naturally arise in the known emergent/analogue spacetimes typically violate analogue Lorentz invariance at high energy, but do not do so in completely arbitrary manner. This suggests that a search for arbitrary violations of Lorentz invariance is possibly overkill: There are a number of natural and physically well-motivated restrictions one can put on emergent/ analogue dispersion relations,  considerably reducing the plausible  parameter space. 

}

\FullConference{From Quantum to Emergent Gravity: Theory and Phenomenology\\
		 June 11-15 2007\\
		 Trieste, Italy}

\begin{document}

\section{Introduction}
\def\x{{\vec{x}}}
\def\d{{\mathrm{d}}}

The word  ``emergence''  is being tossed around an awful lot lately~\cite{emergence, emergent, naturalness}, but without any clear agreement in the quantum gravity community as to what it exactly means.  Various proponents seem to be suggesting one or more of the potential meanings:
\begin{itemize}
\item ``More is different''?~\cite{Anderson}
\item The sum is greater than its parts? 
\item Universality? 
\item  The existence of a ``mean field''?
\end{itemize}
We shall adopt a reasonably conservative (and we think non-controversial) definition that ``emergence'' is equivalent to the observation that:
\begin{itemize}
\item  Short-distance physics is often \emph{radically} different from long-distance physics.
\end{itemize}
In particular, the long-distance physics may sometimes be so radically different from the short-distance physics that it effectively gives no real guide as to what the short-distance physics might look like. It is in this situation that one may say that the long-distance physics is ``emergent''  from the short-distance physics.
A particularly  surprising result is that various analogue models with different microscopic degrees of freedom not only exhibit a geometrical interpretation in the deep infrared limit, they also show ``sensible'' ultraviolet modifications in their collective variables. This increases our trust in using effective field theories without exact knowledge of the full underlying micro-physics --- and therefore, when applied to quantum gravity, increases our trust in quantum gravity phenomenology.
 
Historically and logically the prime example of ``emergence'' is that of fluid dynamics, where the basic equations summarizing the long-distance physics were understood some 100 years before even rudimentary understanding of the short-distance physics was achieved~\cite{Lamb}. Specifically, the key features of this system are:
\begin{itemize}
\item Long-distance physics (fluid mechanics)~\cite{Lamb}:    
\begin{itemize}
\item Euler equation (generic);
\item Continuity equation (generic);
\item Equation of state (specific).
\end{itemize}
\item
Short-distance physics (quantum molecular dynamics):   
\begin{itemize}
\item Schr\"odinger equation (generic);
\item Inter-molecular potential (specific).
\end{itemize}
\end{itemize}
Note that the long-distance physics depends on both generic results (Euler equation, continuity equation), and very specific system-dependent information (the equation of state) that  depends on the particular fluid one is dealing with. Similarly in the short-distance realm one has both fundamental results (Schr\"odinger equation), and ``messy'' system-specific information encoded in the inter-molecular potential.
It is vitally important to realize that you cannot hope to \emph{derive} quantum molecular 
dynamics by quantizing fluid dynamics. Quantum molecular dynamics, and quantum physics itself, was developed using other experimental input, orthogonal to the realm probed by fluid dynamics. (It would be difficult to see how quantum physics might successfully have been developed without input from spectroscopy and the atomic theory  --- both of which fields are quite distinct and quite independent of fluid mechanics itself.)

In the ongoing search for a fully viable theory of ``quantum gravity'', one non-standard but vitally important question is this:  Could Einstein gravity itself be ``emergent''? (And if so, in precisely what sense of the word ``emergent''?) To address this overall question one will need to first focus on two more specific issues:
\begin{itemize}
\item Can we develop a compelling theory that naturally leads to an ``analogue spacetime''?  \\ 
(a generic question) 
\item Can we develop a compelling theory that naturally leads to Einstein's equations? \\ 
(a much more specific question)\footnote{While it is conceivable that there is a reasonably large "generic" class of analogue models that somehow lead to the Einstein equations, it must be emphasized that as yet,  there is not a single known analogue model that cleanly and compellingly leads to the Einstein equations.}
\end{itemize}
There is a  reason this question is of central importance:
\emph{If} Einstein gravity is ``emergent'', (and this is a very big ``if''),
\emph{then} it makes absolutely no sense  to ``quantize gravity'' as a fundamental theory. This observation would then doom many of the currently fashionable approaches to quantum gravity, so it is a good idea to settle this issue as quickly as possible.
If Einstein gravity is emergent, then the best one could hope for is to develop some uber-theory that 
approximately reduces to Einstein gravity in some appropriate limit. 
This need not preclude the use of an ``effective'' quantum field theory [QFT] framework to describe gravity,  but ``emergence'' would strongly suggest that we do not even know what the fundamental degrees of freedom might be, and they would be unlikely to be anything as simple as the metric or vierbein [tetrad]. Indeed, the uber-theory would not even necessarily be quantum (in the usual sense)~\cite{t-Hooft}, though at various levels of approximation it  must exhibit the well-defined and controlled limits: 
\begin{itemize}
\item  Classical Einstein gravity; 
\item  Flat-space (Minkowski) QFT;
\item  Curved space QFT;
\item  Semiclassical quantum gravity (with semiclassical back reaction).
\end{itemize}
The ``analogue spacetimes'' of central importance to this article are (among other things) 
baby steps in this direction.
For example, the simplest  ``analogue spacetimes'' are the ``acoustic spacetimes''~\cite{rimfall}, and there is by now a quite sizable literature on acoustic, and other more general analogue spacetimes~\cite{Unruh, Visser, LRR}. The main message to extract from the literature is this:  Finding an emergent/ effective/ analogue  low-energy metric is not all that difficult, trying to determine or control the dynamics of this low-energy emergent metric is much more difficult.

 A key result underlying the whole analogue spacetime programme, which can perhaps best be viewed as a rigorous \emph{theorem} of mathematical physics, is this~\cite{Unruh,Visser,LRR}:\\
{\bf Theorem:}
Consider a non-relativistic irrotational inviscid barotropic perfect fluid, whose motion is governed by the Euler equation, the continuity equation, and an equation of state $\rho=\rho(p)$.
 Then the dynamics of the linearized perturbations around any background solution of the equations of motion (either static or time dependent) is governed by a D'Alembertian equation 
 \begin{equation}
\Delta_g \Phi \equiv {1\over\sqrt{-g}} \; \partial_a \left( \sqrt{-g} \; g^{ab} \;\partial_b \Phi \right) = 0,
\end{equation}
involving an ``acoustic metric'' $g$ which is an algebraic function of the background fields.  Explicitly, in (3+1) dimensions, we have
\begin{equation}
g_{ab} (t,\x)\equiv 
{\rho_0 \over  c_0}
\begin{bmatrix}
   -(c_0^2-v_0^2)&\vdots&-v_0^j\\
   \cdots\cdots\cdots\cdots&\cdot&\cdots\cdots\\
   -v_0^i&\vdots&\delta_{ij}\\
\end{bmatrix},
\end{equation}
 and
\begin{equation}
g^{ab}(t,\x) \equiv 
{1\over \rho_0 \; c_0}
\begin{bmatrix}
   -1&\vdots&+v_0^j\\
   \cdots\cdots&\cdot&\cdots\cdots\cdots\cdots\\
   +v_0^i&\vdots&(c_0^2 \; \delta^{ij} - v_0^i \; v_0^j )\\
\end{bmatrix}.        
\end{equation}
The linearized perturbations are conventionally referred to as ``sound waves'' or, if quantized, ``phonons'', and $c_0$ is the speed of sound defined by $c_0^2 = \partial p/\partial\rho$.  \hfill $\Box$.

 We emphasize the central importance of this result: It is a rigorous theorem of mathematical physics unambiguously demonstrating the existence of a curved spacetime whose geometry is defined by this acoustic metric --- It is thus a proof of principle that spacetime metrics, and indeed Lorentzian geometries, can naturally ``emerge'' in situations where one would not \emph{a priori} have expected them~\cite{Unruh, Visser, LRR}. 
 
 Of course, the analogue model programme is not limited to acoustic spacetimes --- there has been significant progress in more abstract directions, perhaps closer to the main interests of the ``emergence'' community, based on an extension of the usual notion of ``field theory normal modes''~\cite{normal, birefringent},  and in the other direction there has been much progress in understanding analogue spacetimes in more specific and experimentally realizable systems such as Bose--Einstein condensates~\cite{BEC-basics}.
Indeed by using a  Feschbach resonance to control the scattering length (and hence the speed of sound) in the BEC it is possible to set up analogues of both Friedmann--Robertson--Walker [FRW] universes~\cite{FRW} and more exotic phenomena such as controlled signature-change events~\cite{signature}. In fact, what relativists call a signature change event is from the condensed matter point of view a so-called ``Bose-nova'' event, and signature change physics, with
\begin{equation}
c_0^2    \propto     \hbox{(scattering length)},
\end{equation}        
can be connected, via the interpretation given in~\cite{signature},  to the field theory analysis of mode amplification on a dynamically evolving condensate presented by Calzetta and Hu~\cite{hu-calzetta}.

In short, many interesting  extensions and modifications of the general relativity notion of spacetime have conceptually concrete and well controlled models within the ``analogue spacetime'' framework. 
When developing a ``quantum gravity phenomeniology'' this tells us which rocks to start looking under, and the remainder of this article will deal with a few specific extensions to standard spacetime that can best be motivated, interpreted, and perhaps understood, in terms of the analogue spacetime programme.

\section{Analogue rainbow spacetimes}

We shall now develop several models for  ``rainbow spacetimes'' based on the ``analogue spacetime'' programme. These geometries are particularly useful as concrete conceptual and physical examples of how to construct physically well-motivated rainbow geometries, which may then be of interest as guideposts when considering possible energy-dependent modifications of general relativity. One class of models is based on generalizing the acoustic spacetimes of classical fluid mechanics by inserting the momentum-dependent group velocity and phase velocity into the spacetime metric --- this leads to (at least) two distinct ``rainbow metrics'', which describe distinct aspects of the physics, and which converge on the ordinary acoustic metric in the hydrodynamic limit.\footnote{A second class of energy-dependent geometries, not discussed in this article, can be built by using the Maupertuis form of the least  action principle to rewrite Newton's second law in terms of  geodesic equations on an energy-dependent manifold.} While these particular models are not themselves of direct relevance to quantum gravity, they do provide mathematically and physically well-defined examples of what a ``rainbow spacetime'' should be.

To start the discussion, note that there is no general widely accepted mathematical definition of exactly what is meant by the term  ``rainbow geometry''. The physicist's definition is still rather imprecise, and amounts to one or more of the following: 
\begin{itemize}
\item An ``energy dependent'' metric? 
\item A  ``momentum dependent'' metric? 
\item A  ``4-momentum dependent'' metric? 
\end{itemize}
One particularly acute source of confusion is that it is not clear as to \emph{which} energy/ momentum/ 4-momentum should be used? That of the observer? That of the object being observed?

To capture the essence of ``energy dependence'' we need an object that depends both on the location/coordinate patch \emph{and} at the very least has some dependence on some well-defined notion of energy/momentum. Consider for instance a fluid at rest, in very many cases the dispersion relation can be written in the form\footnote{Making the dispersion relation a function of $k^2$, rather than $\vec k$, implies that we are adopting parity invariance. If one desperately wishes to break parity, go right ahead --- but there seems to be no physically compelling reason to do so. See further discussion below.}
\begin{equation}
\omega^2 = F(k^2)
\end{equation}
for some often nonlinear function $F(k^2)$. This captures the notion of a dispersion relation that is 2nd-order in time (frequency), but arbitrary order in space (wavenumber). Models of this type have particularly been used by Jacobson~\cite{Jacobson} and Unruh~\cite{Unruh-dispersion} to investigate the influence of non-standard dispersion relations on the Hawking radiation from a black hole horizon. 

\subsection{Phase velocity rainbow geometry}
 
In the situation described above one can define a phase velocity by
\begin{equation}
c_\mathrm{phase}^2(k^2) = {\omega^2\over k^2}  = {F(k^2)\over k^2},
\end{equation}
and so without loss of generality rewrite the dispersion relation as
\begin{equation}
\omega^2 = c_\mathrm{phase}^2(k^2) \; k^2.
\end{equation}
For a fluid that is in motion (not necessarily uniform motion) we can simply (non-relativistically) Doppler shift
\begin{equation}
\omega \to \omega - \vec v \cdot \vec k\;;
\qquad
\vec k \to \vec k\;;
\end{equation}
to derive 
\begin{equation}
\left(\omega -\vec{v} \cdot \vec k\right)^2 = c_\mathrm{phase}^2(k^2) \; k^2. 
\end{equation}
Now re-write this dispersion relation (``mass shell condition'') in the form
\begin{equation}
g^{ab}(k^2) \; k_a \; k_b = 0; \qquad  k_a = (\omega; \; \vec k) = (\omega; \; k_i);
\end{equation}
and pick off the coefficients $g^{ab}(k^2)$. We see
\begin{equation}
g^{ab}(k^2) \propto  \left[ \begin{array}{c|c}
\vphantom{\Big|}
- 1 & +v^j \\
\hline
\vphantom{\Big|}
+v^i &  c_\mathrm{phase}^2 (k^2) \; \delta^{ij} - v^i \; v^j \\
\end{array}
\right].
\end{equation}
which we interpret as a (contra-variant) ``inverse metric'', which is 3-momentum dependent because of its dependence (through the phase velocity) on the wavenumber $k$. Performing a matrix inversion, we see that the related  (co-variant) ``metric'' is\footnote{While the phase velocity $c_\mathrm{phase}(k^2)$ is defined as a function of momentum, it nevertheless provides crucial information about about how wavefronts move in physical configuration space. Recall that the \emph{phase} of a plane wave is given by $\vec k \cdot \vec x - \omega \; t=  \vec k \cdot ( \vec x - c_\mathrm{phase} \, \hat n \, t)$.  Thus phase velocity gives you information about the physical location of interference fringes, and the like, in the physical configuration space.}
\begin{equation}
g_{ab}(k^2) \propto  \left[ \begin{array}{c|c}
\vphantom{\Big|}
- \left\{c_\mathrm{phase}^2(k^2) - \delta_{ij} \; v^i\; v^j \right\} & -v_j \\
\hline
\vphantom{\Big|}
-v_i &  + h_{ij}  \\
\end{array}
\right].
\end{equation}
All in all, this now gives a precise and well-motivated mathematical definition of a rainbow metric depending on the phase velocity.  We also note that in this particular approach the inverse metric $g^{ab}(k^2)$ is more ``primitive'' (more basic) than the metric itself, since it is the inverse metric that directly shows up in the dispersion relation.

 A particularly nice feature of this dispersion relation approach is that it is physically transparent.
The only real weakness likes in the occurrence of proportionality signs instead of equalities:   the conformal factor is left unspecified in this approach and the rainbow geometry we have written down is really a conformal class of geometries. This is a quite standard side-effect of making the geometrical quasi-particle approximation, and looking at the mass shell condition (dispersion relation)  of the excitations. While we have here phrased the discussion in terms of geometrical acoustics, the same conformal ambiguity arises in  geometrical optics, or more generally in any situation where a partial differential equation is approximated by an \emph{eikonal}.  Only if one has direct access to the underlying PDE, (\emph{e.g.}, the hydrodynamic fluid equations), which contains more information than the dispersion relation itself, does one have a hope of specifying the overall conformal factor.\footnote{This happens, for instance, in Bose--Einstein condensates driven beyond the hydrodynamic limit, where one not only has the Bogoluibov dispersion relation as discussed below, but because one additionally has direct access to the microphysics in terms of the linearized Gross--Pitaevskii equation,  the overall conformal factor can be uniquely defined in a physically compelling manner~\cite{FRW,signature}.}  Note in particular that the momentum in question is now clearly and unambiguously the momentum 
of an individual  ``mode'' of the field.

\subsection{Group velocity rainbow geometry}

Similar but distinct steps can now be taken to develop a notion of rainbow spacetime  based on the group velocity.  Consider a ``wave packet'' centered on 3-momentum $p=\hbar\; k$, that is centered on 3-wavevector $\vec k$. Such a wave packet will (essentially by definition) propagate at the group velocity 
\begin{equation}
c_\mathrm{group}(k^2) = {\partial \omega\over \partial k}  = {\partial\sqrt{F(k^2)}\over \partial k},
\end{equation}
but note that the packet moves at this speed with respect to the background velocity of the fluid. That is, if the wave packet moves a distance $\d \vec x$ in a time $\d t$ then these increments must satisfy
\begin{equation}
\left(\d \vec x - \vec v \; \d t\right)^2 = c_\mathrm{group}^2(k^2) \; \d t^2.
\end{equation}
Now rewrite this as a ``sound cone'' condition
\begin{equation}
\d s^2 = 0 =  g_{ab}(k^2) \; \d x^a \; \d x^b;  \qquad  \d x^a = (\d t; \; \d\vec x) = (\d t; \; \d x^i),
\end{equation}
and pick off the coefficients of the (co-variant) ``group velocity metric'' $g_{ab}(k^2)$. We see
\begin{equation}
g_{ab}(k^2) \propto  \left[ \begin{array}{c|c}
\vphantom{\Big|}
- \left\{c_\mathrm{group}^2(k^2) - \delta_{ij} \; v^i\; v^j \right\} & -v_j \\
\hline
\vphantom{\Big|}
-v_i &  + h_{ij}  \\
\end{array}
\right],
\end{equation}
while for the (contra-variant) inverse metric we obtain
\begin{equation}
g^{ab}(k^2) \propto  \left[ \begin{array}{c|c}
\vphantom{\Big|}
- 1 & +v^j \\
\hline
\vphantom{\Big|}
+v^i &  c_\mathrm{group}^2 (k^2) \; \delta^{ij} - v^i \; v^j \\
\end{array}
\right].
\end{equation}
This now gives a precise and well-motivated mathematical definition of a rainbow metric depending on the group velocity.  We also note that in this particular  approach the metric $g_{ab}(k^2)$ is more ``primitive'' (more basic) than the inverse metric, since it is the metric that directly shows up in the sound cone condition.
Again there is an undetermined conformal factor, now due to the fact that the sound cone condition is conformally invariant.\footnote{This is quite standard and exactly the same thing happens in standard general relativity: If one only has the light cones (the causal structure) then the spacetime metric is determined only up to overall local conformal deformation.} Only with additional information, external to what one can extract from the dispersion relation, would one have any hope of meaningfully specifying this conformal factor. Note in particular that the momentum in question is now clearly and unambiguously the momentum 
of the wave packet.

\subsection{More general rainbow geometries}

So we have seen that there are \emph{at least} two distinct and very different notions of  ``rainbow spacetime'' in an analogue setting. The phase velocity and group velocity geometries answer different physical questions:
\begin{itemize}
\item Phase velocity: What is the dispersion relation for a pure mode?
\item Group velocity: How does a wave packet propagate?
\end{itemize}
If one is \emph{lucky} these two geometries will converge on a low-momentum ``hydrodynamic limit''. This occurs if 
\begin{equation}
\lim_{k\to 0} c_\mathrm{phase}^2(k^2) = c_\mathrm{hydrodynamic}^2 
=  \lim_{k\to 0} c_\mathrm{group}^2(k^2),
\end{equation}
and if this occurs the common limit is more typically referred to as ``the'' speed of sound $c_0$.
Note that there is nothing in principle to stop us from identifying other physically relevant notions of ``speed'' and using them to construct more general rainbow geometries:
\begin{equation}
g_{ab}(k^2) \propto  \left[ \begin{array}{c|c}
\vphantom{\Big|}
- \left\{c^2(k^2) - \delta_{ij} \; v^i\; v^j \right\} & +v_j \\
\hline
\vphantom{\Big|}
+v_i &  + h^{ij}  \\
\end{array}
\right].
\end{equation}
\begin{equation}
g^{ab}(k^2) \propto  \left[ \begin{array}{c|c}
\vphantom{\Big|}
- 1 & +v^j \\
\hline
\vphantom{\Big|}
+v^i &  c^2 (k^2) \; \delta^{ij} - v^i \; v^j \\
\end{array}
\right].
\end{equation}
At a minimum we could think of using the following notions of propagation speed\footnote{Brillouin, in his classic reference~\cite{Brillouin}, identified at least six useful notions of propagation speed, and many would argue that the list can be further refined.}
\begin{equation}
c(k^2) \to \left\{ \begin{array}{l}
c_\mathrm{phase}(k^2);\\
c_\mathrm{group}(k^2);\\
c_\mathrm{hydrodynamic};\\
c_\mathrm{signal};\\
\infty?\\
\end{array}
\right. 
\end{equation}
The most standard definition of signal velocity is
\begin{equation}
c_\mathrm{signal,1} = \lim_{k\to\infty} c_\mathrm{phase}(k^2),
\end{equation}
which focusses on the mathematically precise definition of how discontinuities propagate. A perhaps more operational definition
\begin{equation}
c_\mathrm{signal,2} = \max_{k} \; c_\mathrm{group}(k^2),
\end{equation}
focusses on the more physical question of how rapidly one can transfer information encoded in wave-packets. 

It is a deep issue of principle that as long as the signal velocity is finite the overall causal structure will be similar to that of general relativity,  just with ``signal cones'' instead of light cones.
If however the signal velocity is infinite then the global overall structure will be similar to that of Newtonian physics.   The more traditional distinction between ``superluminal'' and ``subluminal'' dispersion relations,  while it certainly impacts on particle scattering and particle production thresholds,   and so greatly constrains allowable particle interactions,  is very much of subsidiary importance 
when it comes to determining overall causal structure.

\section{A direct connection with quantum gravity phenomenology}

To connect these ``analogue spacetimes'' back to quantum gravity phenomenology, consider the following argument.  Possible ultra-high energy violations of Lorentz invariance are one of the most important  ``signals'' being considered when developing quantum gravity phenomenology. The basic idea is this:  Consider a standard Lorentz-invariant dispersion relation
\begin{equation}
\omega^2 = \omega_0^2 + c^2 \; k^2,
\end{equation}
which we might wish to replace with something such as
\begin{equation}
F_1(\omega,k) = 0,
\end{equation}
or, after appealing to the implicit function theorem
\begin{equation}
\omega = F_2(k).
\end{equation}
But a completely general dispersion relation of this type is too general to actually be particularly useful,  and it is prudent to restrict one's attention in suitable manner. Let us introduce a number of ``working hypotheses'' to see where they lead us.

\subsection{Hypothesis 1: Not just $CPT$, but $C$, $P$, and $T$.}
Classical Einstein gravity is certainly invariant under charge conjugation, parity inversion, and time reversal --- Einstein gravity is $C$,  $P$,  and  $T$  invariant.  Furthermore, physically the only known examples of  $P$ and $T$ violations are in the  electro-weak sector,  seemingly unconnected 
with gravity.  Certainly no gravitational experiment has ever detected $P$ or $T$  violations. So don't add more complications than necessary. Working hypothesis 1 will be that in developing quantum gravity phenomenology we maintain both $P$ and $T$ invariance unless and until we absolutely have to sacrifice these these symmetries.\footnote{This is in some sense the theorists' version of the well-known experimentalists' adage "Only change one variable at a time". When moving beyond the known and established theoretical framework it is best to minimize the number of radical steps one invokes. If nothing else, this will help the theorist isolate which exotic effect in his/her toy model is due to which exotic cause.  We are certainly \emph{not} claiming that the ultimate theory of quantum gravity is   guaranteed to be $C$,  $P$,  and  $T$  invariant, just that given the known fact that Einstein gravity is  $C$,  $P$,  and  $T$  invariant, it is not particularly likely that violations of  $C$,  $P$,  and/ or  $T$ would be the first most obvious signal of quantum gravity phenomenology.} Adopting these symmetries, a minimal basis of terms invariant under $P$ and $T$ is:
\begin{equation}
\omega^2;  \qquad \omega\, (\vec v \cdot \vec k); \qquad (\vec v \cdot \vec k)^2;  \qquad h^{ij}\;k_i\; k_j. 
\end{equation}
Here $\vec v(\vec x, t)$, which can in principle depend on space and time, is some quantity which transforms \emph{like} a velocity under $P$ and $T$, we do not yet claim that it \emph{is} a velocity.
Similarly $h^{ij}(\vec x, t)$ transforms under $P$ and $T$ \emph{like} a $3\times3$ matrix of dielectric constants, but is otherwise unconstrained. One could in addition think of including 4th-order terms such as $h^{ijkl} \;k_i\; k_j\;k_k\; k_l$, but we shall exclude these on the grounds that they would not be ``minimal''.\footnote{Adding terms of this type would lead to a Finsler, or more precisely, pseudo--Finsler dispersion relation, which may ultimately be of interest for other reasons --- but not just now~\cite{LNP}.}
Adopting this $P$ and $T$ invariant basis, the general dispersion relation $F_2$ can be cast in the form
\begin{equation}
F_3\left(\;\omega^2;\;  \omega\, (\vec v \cdot \vec k);\;   (\vec v \cdot \vec k)^2; \; h^{ij}\;k_i\; k_j\;\right) = 0. 
\end{equation}
Without loss of generality we can recombine the terms in the $P$ and $T$ invariant basis to obtain 
\begin{equation}
\omega^2;  \qquad (\omega -\vec v \cdot \vec k)^2; 
\qquad (\vec v \cdot \vec k)^2;  \qquad  h^{ij}\;k_i\; k_j;
\end{equation}
which then lets us cast the dispersion relation in the alternative form
\begin{equation}
F_4\left( \, \omega^2;\:  (\omega -\vec v \cdot \vec k)^2; 
\; (\vec v \cdot \vec k)^2;  \;h^{ij}\;k_i\; k_j\; \right)=0. 
\end{equation}

\subsection{Hypothesis 2: Time-derivatives higher than 2nd order tend to be problematic.}

Time derivatives higher than second-order tend to lead to ghosts and unitarity violations, and are 
rarely seen in nature.  The only significant exception are the Fresnel relations in optics. But in (almost) all known physically relevant cases,  Fresnel relations factorize into second-order fragments. Consider a typical Fresnel relation in (3+1) dimensions: 
\begin{equation}
\omega^2\;(A \,\omega^4 + B\,\omega^2 \,k^2 + C \,k^4) =0.
\end{equation}
This includes the effects of two physical ``transverse'' photon polarizations, plus one unphysical ``longitudinal'' polarization. But in most known cases (including uni-axial bi-refringent crystals) the Fresnel relation factorizes
\begin{equation}
\omega^2 \;(\omega^2-c_1^2 \, k^2)\; (\omega^2 - c_2^2 \, k^2) = 0,
\end{equation}
leading to so-called ``ordinary'' and ``extra-ordinary'' rays. The situation for bi-axial bi-refringent crystals is considerably worse: The Fresnel relation does not then factorize, and the geometry is actually then of Finsler type (more precisely, pseudo--Finsler). Nevertheless, it is clear that in almost all situations the occurrence at most of 2nd order time derivatives is preferred\footnote{Infinite-order time derivatives typically (but not quite always) lead to non-locality in time, and are typically associated with causality problems.} so that we can plausibly write our quantum gravity inspired dispersion relation as:   
\begin{equation}
F_5\left( \, a \,\omega^2 + b \, (\omega -\vec v \cdot \vec k)^2; \;   (\vec v \cdot \vec k)^2; \;  h^{ij}\;k_i\; k_j\; \right)=0. 
\end{equation}
Combining the two quadratic in frequency terms
\begin{equation}
 a \,\omega^2 + b \, (\omega -  \vec v \cdot \vec k)^2 = 
 [a+b] \left\{  \left(\omega -\vec {\bar v} \cdot \vec k\right)^2 
 + {a\over b} \,  \left(\vec {\bar v} \cdot \vec k\right)^2 \right\},
 \end{equation}
 where
 \begin{equation}
 \vec{\bar v} = {b\over a+b} \; \vec v,
 \end{equation}
 now permits us to write:
\begin{equation}
F_6\left(  \left(\omega -\vec{\bar v} \cdot \vec k\right)^2; \;  (\vec{\bar  v} \cdot \vec k)^2; \;  h^{ij}\;k_i\; k_j\; \right)=0. 
\end{equation}
Appealing to the implicit function theorem this becomes
\begin{equation}
\left(\omega -\vec{\bar v} \cdot \vec k\right)^2 = F_7\left( \; h^{ij}\;k_i\; k_j;  \; (\vec {\bar v} \cdot \vec k)^2\; \right). 
\end{equation}
Finally, drop unnecessary subscripts and over-bars:
\begin{equation}
\left(\omega -\vec{v} \cdot \vec k\right)^2 = F\left( \; h^{ij}\;k_i\; k_j;\; (\vec v \cdot \vec k)^2\; \right). 
\end{equation}
That is, under two very mild conditions ($P$ and $T$ invariance, and limiting time derivatives to 2nd order)  we can plausibly write quantum gravity phenomenology inspired dispersion relations as
\begin{equation}
\label{E:H12}
\left(\omega -\vec{v} \cdot \vec k\right) = \sqrt{F\left( \; h^{ij}\;k_i\; k_j;\; (\vec v \cdot \vec k)^2\; \right)}. 
\end{equation}
But such dispersion relations fall naturally into a minor extension of the class of dispersion relations arising naturally in ``emergent/ analogue'' spacetimes. As we have seen in the previous section, based on ``analogue spacetimes'' it is natural to expect
\begin{equation}
\left(\omega -\vec{v} \cdot \vec k\right) = \sqrt{\tilde F\left( \; \delta^{ij}\;k_i\; k_j \;\right)}. 
\end{equation}
Note however that to get to equation (\ref{E:H12}) we have not used any ``analogue model'' reasoning, just some very mild and fundamental working hypotheses --- what would seem to be eminently reasonable constraints that (most) quantum gravity phenomenologies should satisfy. 
Note in particular that no notion of Lorentz invariance has been used, nor should it be used since we are in particular interested in looking for deviations from ordinary Lorentz invariance. Let us now see how much further we can go.

\subsection{Hypothesis 3: Taking $\vec v$ seriously.}

Now let's take the quantity $\vec v$ a little more seriously, and hypothesize that it really is some 
sort of   ``physical velocity''. For instance, the local preferred rest frame for Lorentz breaking? 
Or,  (if you like to give colleagues heart attacks), you could call this the ``velocity of the sub-quantic aether''. Whatever you choose to call it does not matter:   If it is a physical velocity,  then you can certainly go to the local rest frame where this velocity is zero. Doing so reduces the dispersion relation $F_7$ to
\begin{equation}
\omega^2 = F_8\left( \; h^{ij}\;k_i\; k_j \right). 
\end{equation}
Let us emphasize that we are not using Lorentz transformations to effect this reduction, in fact we have no idea what suitable symmetry transformations might be, or even if there are suitable symmetry transformations. All that is going in is this: if $\vec v$ really is a physical velocity then one should (at least locally) be able to move at velocity $\vec v$, effectively putting you ``at rest'' with respect to whatever it is that $\vec v$ is representing.
In the local rest frame, dropping unnecessary subscripts
\begin{equation}
\omega^2 = F\left( \; h^{ij}\;k_i\; k_j \right). 
\end{equation}
Defining the phase velocity by
\begin{equation}
c_\mathrm{phase}^2(k^2) = {F\left( \; h^{ij}\;k_i\; k_j \right)\over h^{ij}\;k_i\; k_j},
\end{equation}
the dispersion relation becomes\footnote{Indeed one could get directly from $F_4$ to (\ref{E:F_phase}) by ignoring hypothesis 2 and directly appealing to hypothesis 3. The reason we do not do so is because hypothesis 3 is in many ways considerably stronger than hypothesis 2, and it is useful to see how far one can go with the weaker hypothesis.}
\begin{equation}
\label{E:F_phase}
\omega^2 = c_\mathrm{phase}^2(k^2) \;\left\{ \; h^{ij}\;k_i\; k_j \; \right\}. 
\end{equation}
Again, note the very strong similarities between analogue model inspired dispersion relations and quantum gravity phenomenology inspired dispersion relations.

\subsection{Hypothesis 4: For some purposes, Galileo rules.}

For the final step, no-one can stop us from making a Galilean coordinate transformation
\begin{equation}
\vec x \to \vec x + \vec v \; t;
\qquad
t \to t;
\end{equation}
or more precisely, the local infinitesimal version
\begin{equation}
\d \vec x \to \d\vec x + \vec v(\vec x,t) \; \d t;
\qquad
\d t \to \d t.
\end{equation}
We make no claim that this is in any sense a symmetry of the system, it is ``merely'' a convenient choice of coordinates. Of course a change in spacetime coordinates induces a change in the co-tangent space coordinates as well:
\begin{equation}
\omega \to \omega - \vec v \cdot \vec k\;;
\qquad
\vec k \to \vec k\;.
\end{equation}
This now implies
\begin{equation}
\left(\omega -\vec{v} \cdot \vec k\right)^2 = F\left( \; h^{ij}\;k_i\; k_j\; \right).
\end{equation}
or equivalently
\begin{equation}
\left(\omega -\vec{v} \cdot \vec k\right)^2 = c_\mathrm{phase}(k^2) \; \left\{ \; h^{ij}\;k_i\; k_j\; \right\}.
\end{equation}
This finally gives us a wide class of quantum gravity phenomenology inspired dispersion relations that are formally identical (up to the formal replacement $h^{ij} \leftrightarrow \delta^{ij}$)  with the analogue model inspired dispersion relations.

\subsection{Comments}
The arguments just presented give us confidence that whatever  insights we extract from the ``emergent analogue spacetime'' programme are likely to be generic  to a wide class of physically reasonable  quantum gravity phenomenologies. 
Note that we are not asserting to have demonstrated that all quantum gravity phenomenology inspired dispersion relations must be of analogue spacetime form --- just the milder statement that it is ``natural'' that they are of analogue spacetime form. 
In particular, if one wants to move beyond the class of dispersion relation naturally arising in analogue models, this analysis can be inverted to delineate just what is required to do so. (For example, adding the quartic parity invariant $h^{ijkl} \;k_i\; k_j\;k_k\; k_l$ will naturally lead to a Finsler structure; adding explicit uncontrolled $P$ or $T$ violations will lead to an extremely general and essentially useless dispersion relation, \emph{etc.})\footnote{There are for instance systems based on liquid He3 that explicitly break parity~\cite{droplet}. One can then argue as to how generic this behaviour is in the set of all analogue spacetimes, and whether there is any need to introduce explicit parity breaking given the known phenomenology of classical gravity.}

\section{Some specific analogue dispersion relations}

Now that we have seen some of the general features that analogue model inspired dispersion relations can have, and the close way in which they track quantum gravity phenomenology inspired dispersion relations, it is time to look at some specific analogue examples, and consider what they might tell us.

\subsection{Bogoliubov spectrum}
The Bogoliubov dispersion relation
\begin{equation}
\omega^2 = c_0^2\; k^2 + {\hbar^2 \; k^4\over (2 m)^2},
\end{equation}
is of interest in both BECs and superconductors. In a BEC context $m$ represents the mass of the atoms undergoing condensation, while in a superconductor context $m$ represents the mass of the Cooper pairs. Note that at low wavenumber the quasiparticle is ``relativistic'', $\omega \sim c_0 \, k$, while at large wavenumber  it is ``Newtonian'', $\omega\sim \hbar^2 k^2/(2m)$.  Rewriting the dispersion relation as 
\begin{equation}
\omega^2 = c_0^2\; k^2 \left\{ 1 + k^2/K^2\right\},
\end{equation}
it is easy to see that
\begin{equation}
c_\mathrm{phase} = {\omega\over k} = c_0  \; \sqrt{1 + k^2/K^2} \geq c_0;
\end{equation}
\begin{equation}
c_\mathrm{group} = {\partial\omega\over \partial k} 
= c_0  \; {1+ 2 k^2/K^2\over \sqrt{1 + k^2/K^2}} \geq c_0.
\end{equation}
In particular, since for nonzero wavenumber $c_\mathrm{group} > c_0$, this dispersion relation is said to be ``supersonic'', and provides an analogue for the ``superluminal'' dispersion relations often encountered in quantum gravity phenomenology.\footnote{Note that ``superluminal'' and ``subluminal'' \emph{must} be defined in terms of group velocity,  attempting to work with phase velocity gives meaningless answers even for standard special relativity.} Furthermore $K$ can be interpreted as the scale of ``Lorentz symmetry breaking'' and this dispersion relation provides a simple and controlled example of Lorentz symmetry breaking~\cite{broken}.\footnote{Indeed, preserving Lorentz invariance at high energies is in this context a fine-tuning issue. The generic behaviour is that  one encounters low-energy ``Lorentz invariance'' coupled with high-energy breaking of the ``Lorentz invariance''. Suppressing the high-energy Lorentz breaking then requires extra symmetry (possibly via bose-fermi cancellations) and/ or a fine-tuning of counter-terms.} By looking at large wavenumbers it is easy to see that the signal velocity is infinite $c_\mathrm{signal}=\infty$, so that the overall causal structure is Galilean. Finally note that the wave and group velocites never differ by more than a factor of 2:
\begin{equation}
{c_\mathrm{group}\over c_\mathrm{phase} }
=  {1+ 2 k^2/K^2\over 1 + k^2/K^2}  =   {K^2 + 2 k^2\over K^2 + k^2} \in [1,2].
\end{equation}
Turning to more general situations, by letting modes interact with each other,  it is possible to introduce quasiparticle masses~\cite{quasiparticle-mass}
\begin{equation}
\omega^2 =  \omega_0^2 + c_0^2\; k^2 + c_0^2 \;{k^4\over K^2} + {O}[k^6] .
\end{equation}
The mass-generating mechanism is in this situation intimately related to the Lorentz symmetry breaking mechanism, leading to a natural suppression of low-dimension Lorentz symmetry breaking operators~\cite{naturalness,LNP}.

\subsection{Surface waves in water}

Surface waves in water provide a number of simple analogue models for interesting dispersion relations, that we will now investigate in some detail.

\subsubsection{Shallow-water surface waves with surface tension}
Consider shallow water surface waves, including the effects of surface tension. The relevant
PDE is well-known~\cite{Lamb}
\begin{equation}
\ddot h = g \,d \, \nabla^2 h - {\sigma \, d\, (\nabla^2)^2 h\over \rho}.
\end{equation}
Here $\sigma$ represents the surface tension, $d$ the resting depth of water, $g$ the acceleration due to gravity, and $\rho$ is the density.  Note that the wavelength $\lambda$ cannot be too small or the shallow water approximation breaks down. That is, we want
\begin{equation}
\lambda \gg d.
\end{equation}
Furthermore the transition from gravity-dominated to surface-tension-dominated occurs for
\begin{equation}
\lambda \sim \sqrt{\sigma\over\rho \, g} \equiv K_\sigma^{-1}.
\end{equation}
For water under normal conditions ($\sigma =  72$ dynes/cm; $\rho = 1$ gm/cc; $g = 980$ cm/s$^2$) one has
\begin{equation}
K_\sigma^{-1} \equiv \sqrt{\sigma\over\rho \, g} = 0.27 \hbox{ cm}.
\end{equation}
If we want the shallow water approximation to hold well into the surface-tension-dominated regime we must demand
\begin{equation}
d \ll 0.27 \hbox{ cm},
\end{equation}
so that we are limited to looking at ripples in an extremely thin sheet of water.
Under these conditions we derive the dispersion relation
\begin{equation}
\omega^2 = g\, d\, k^2 + {\sigma\, d\, k^4\over\rho }.
\end{equation}
Formally, this is exactly of the Bogoliubov form with the substitutions
\begin{equation}
c_0 \leftrightarrow \sqrt{g \, d}; \qquad   {\sigma \,d\over\rho} \leftrightarrow {\hbar^2\over(2m)^2};
\qquad
K = {2 m c_0\over \hbar} \leftrightarrow \sqrt{g\;\rho\over\sigma} = K_\sigma^{-1}.
\end{equation}
But note the limited range of validity: This dispersion relation only has interesting ``switchover'' behaviour if
\begin{equation}
d \ll  \sqrt{\sigma\over\rho \, g};    \qquad  K_\sigma d \ll 1; \qquad \hbox{that is} \qquad
d \ll 0.27 \hbox{ cm}.
\end{equation}
Apart from this limitation on the parameters, the dispersion relation is  of Bogoliubov type and provides a second route to ``supersonic'' dispersion relations, quite similar to that for BECs and/or superconductors.

\subsubsection{Shallow-water surface waves (without surface tension)}
In shallow water one has the well-known and very standard result~\cite{Lamb} that surface waves propagate with a frequency-independent constant speed
\begin{equation}
c_0 = \sqrt{ g \; d};       \qquad \lambda \gg d.
\end{equation}
This observation was then used by Sch\"utzhold and Unruh as the basis for their ``shallow water'' analogue for curved spacetime~\cite{shallow}.

\subsubsection{Deep-water surface waves (without surface tension)}
In contrast, in deep water one has the (at first sight unexpected) result that~\cite{Lamb}:
\begin{equation}
\omega = \sqrt{g \; k};  \qquad  \omega^2 = g k;  \qquad \lambda \ll d.
\end{equation}
Naively, the occurrence of an odd power of $k$ seems to violate parity invariance, but we shall soon see the reason for this odd behaviour.  The phase and group velocities are rather odd:
\begin{equation}
c_\mathrm{phase} =  {\omega\over k} = \sqrt{g/k};
\end{equation}
\begin{equation}
c_\mathrm{group} = {\partial\omega\over \partial k} 
= {\sqrt{g/k}\over 2} = {c_\mathrm{phase}\over2}.
\end{equation}
Note that in this situation there is no hydrodynamic limit, as the phase and group velocities \emph{never} converge to each other.  We shall soon see the reason for this odd behaviour. 
Also observe that long wavelengths travel faster, and that for the two at first sight natural definitions of signal velocity we have
\begin{equation}
\lim_{k\to\infty} c_\mathrm{phase} = 0,
\end{equation}
while
\begin{equation}
\max_{k} \; c_\mathrm{group} = \infty,
\end{equation}
with the maximum occurring a $k=0$. So the two natural definitions of signal velocity differ rather drastically. Overall, this situation does not provide a useful analogue for the purposes of quantum gravity phenomenology. (This is also a useful lesson --- it is simply not true that any random condensed matter dispersion relation is automatically interesting for quantum gravity phenomenology.)

\subsubsection{Finite-depth surface waves (without surface tension)}

Consider now surface water waves in an ocean of finite depth. The analysis is again quite standard, see specifically the derivation of the dispersion relation in Lamb~\cite{Lamb} \S 228, p354, eq (5):
\begin{equation}
\omega = \sqrt{g \,k \,\tanh(k\,d)}.
\end{equation}
It is useful to define
\begin{equation}
c_0^2 = g\; d,
\end{equation}
which we shall soon see is the propagation speed in the ``hydrodynamic limit'', and so rewrite the dispersion relation as 
\begin{equation}
\omega^2 = g \, k\, \tanh(k\,d) 
= c_0^2\; k^2 \; {\tanh(k\,d)\over k\, d},
\end{equation}
so that the phase velocity becomes
\begin{equation}
c^2_\mathrm{phase} =  c_0^2\; k^2 \; {\tanh(k\,d)\over k\, d}.
\end{equation}
As initially shallow water begins to get deeper a Taylor series expansion shows
\begin{equation} 
\omega^2 = c_0^2\; k^2 \; \left\{ 1 - {(k\,d)^2\over 3} +      {2(k\,d)^2\over 15} +     \dots  \right\}
\end{equation}
so the first correction is a ``wrong-sign Bogoliubov-like piece''. Indeed for the phase velocity one has
\begin{equation}
c_\mathrm{phase} =  {\omega\over k} = \sqrt{g \,\tanh(k\, d)\over k} = 
c_0 \;  \sqrt{\tanh(k\, d)\over k d } \leq c_0,
\end{equation}
so that for nonzero wavenumber the phase velocity is always less than $c_0$.  The equivalent formula for group velocity is rather ugly  [\emph{cf.} Lamb \S 236, p381] 
\begin{equation}
c_\mathrm{group} = {\partial\omega\over \partial k} 
= c_0 \;  \sqrt{\tanh(k\, d)\over k d } \; \left\{{1\over2} + {k d\over\sinh(2kd)}\right\} \leq c_0.
\end{equation}
Note that for nonzero wavenumber the group velocity is always less than $c_0$, and that both the phase and group velocities converge to $c_0$ as the wavenumber approaches zero. This justifies the identification of $c_0$ as the hydrodynamic propagation speed, and also  demonstrates that this dispersion relation is ``subsonic'',  and so a good model for the ``subluminal'' dispersion relations often encountered in quantum gravity phenomenology.
Note that the group and phase velocities never differ by more than a factor 2:
\begin{equation}
{c_\mathrm{group}\over c_\mathrm{phase} }
= {1\over2} + {k d\over\sinh(2kd)} \in \left[{1\over2},1\right].
\end{equation}
Furthermore
\begin{equation}
\lim_{k\to\infty} c_\mathrm{phase} = 0,
\end{equation}
while on the other hand
\begin{equation}
\max_{k} \; c_\mathrm{group} = c_0^2.
\end{equation}

The preceding discussion also clarifies what was so odd about the infinite depth limit. The true dispersion relation is
\begin{equation}
\omega^2 = g \, k\, \tanh(k\,d) 
= c_0^2\; k^2 \; {\tanh(k\,d)\over k\, d},
\end{equation}
which clearly is $P$ invariant ($k \leftrightarrow - k$). The $d\to\infty$ limit, with its \emph{apparent} $P$ violation, is then rather formal and unphysical. (Infinitely deep oceans simply do not exist, and in particular violate the flat earth approximation that underlies all these dispersion relations) .

Turning to quantum gravity phenomenology, the discussion above suggests a strategy for obtaining an \emph{effective} $k^3$ term in the dispersion relation, but without any fundamental breaking of $P$ invariance. Postulate a dispersion relation of the form
\begin{equation}
\omega^2 = \omega_0^2 + c_0^2\; k^2 +  c_0^2 \; (k^4/K_1^2) \; {\tanh(k/K_2)\over (k/K_2) }.
\end{equation}
Of the two scales occurring here $K_1$ is associated with the breakdown of Lorentz symmetry, while $K_2$ is associated with \emph{apparent} $P$ violation. If the apparent $P$ violation is to be of phenomenological interest, it should kick in before one reaches the Lorentz breaking scale, implying  $K_2 \ll K_1$. Since we normally expect the Lorentz breaking scale to be at the Planck scale, this further implies that if one wishes to include significant $k^3$ terms then even in the pure gravity sector parity must (either effectively as above, or explicitly as in some other models) be broken at sub-Planckian scales.

\subsubsection{Deep-water surface waves with surface tension}

Consider now an infinitely deep ocean, with surface tension, and a free surface. See for instance
Lamb~\cite{Lamb} \S 266, p 459. The dispersion relation is well-known:
\begin{equation}
\omega^2 =  g k +  {\sigma k^3\over\rho},
\end{equation}
so that
\begin{equation}
c^2_\mathrm{phase} = {g\over k} + {\sigma k\over\rho}.
\end{equation}
The phase velocity has a minimum at
\begin{equation}
k^2 = {g\rho\over\sigma} = K_\sigma^2,
\end{equation}
and at this minimum
\begin{equation}
c^2_\mathrm{phase} = 2 \sqrt{g \sigma\over\rho}.
\end{equation}
The group velocity is
\begin{equation}
c_\mathrm{group} = {\partial\omega\over\partial k} = {\partial\sqrt{gk + \sigma k^3/\rho}\over\partial k}
={g+3\sigma k^2/\rho\over2\sqrt{gk + \sigma k^3/\rho}}.
\end{equation}
There is no hydrodynamic limit at low wavenumber and the best one can say is
\begin{equation}
{ c_\mathrm{group} \over c_\mathrm{phase} } ={g+3\sigma k^2/\rho\over2(g + \sigma k^2/\rho)}
\in \left[{1\over2},{3\over2}\right]. 
\end{equation}
The signal velocity, with either definition, is now infinite. All in all, 
this does not provide a useful analogue for the purposes of quantum gravity phenomenology, but serves as a stepping stone to the next dispersion relation we shall consider. 

\subsubsection{Finite-depth surface waves with surface tension}
Consider now surface waves on a finite depth ocean, with surface tension, and a free surface. This is now \emph{not} simply a matter of quoting a well-known result, and it requires a brief computation (based on Lamb~\cite{Lamb} \S 267, p 459) to establish the dispersion relation:
\begin{equation}
\omega^2 =  c_0^2\; k^2\; 
\left\{ 1 + {k^2\over K_\sigma^2}  \right\} \; {\tanh(kd)\over kd}.
\end{equation}
Equivalently the phase velocity is
\begin{equation}
c^2_\mathrm{phase} =  c_0^2\;
\left\{ 1 + {k^2\over K_\sigma^2} \right\} \; {\tanh(kd)\over kd}.
\end{equation}
All symbols have the same meaning as in the previous discussion.

\paragraph{Derivation:} Adapting the discussion in Lamb~\cite{Lamb} \S 267, p 459, we present a brief sketch of the derivation. (A full derivation will almost certainly be available somewhere in the fluid mechanics literature, but an analysis of this particular problem does not appear to be easy to find.) Consider a wave propagating in the $x$ direction, and translationally uniform in the $z$ direction. The positive $y$ direction will be taken to point upwards.  Based on general symmetry principles and the wave equation, the  velocity potential takes the form 
\begin{equation}
\phi(x,y,t) = C \cosh(k[y+d]) \; \cos(kx) \; \cos(\omega t+\epsilon).
\end{equation}
Similarly the height of the free surface takes the form
\begin{equation}
\eta(x,t) =  a\;  \cos(kx) \; \sin(\omega t+\epsilon).
\end{equation}
These two functions are connected at the surface by the boundary condition, [essentially a definition],
\begin{equation}
\dot\eta = - \partial_y \phi \qquad \hbox{(at surface)},
\end{equation}
whence we can relate some of the otherwise arbitrary parameters  
\begin{equation}
\omega \; a = - C \; k \; \sinh(kd).
\end{equation}
But everywhere in the bulk of the fluid the pressure is determined by
\begin{equation}
{p\over\rho} = \dot \phi - g y \qquad\qquad \hbox{(in bulk)},
\end{equation}
and in particular at the free surface we have
\begin{equation}
{p\over\rho} = \dot \phi - g \eta  \qquad \hbox{(at surface)}.
\end{equation}
But the pressure at the surface is also related to the curvature of the surface ($\partial_x^2\eta$) via the surface tension
\begin{equation}
p = - \sigma \partial_x^2 \eta \qquad \hbox{(at surface)}.
\end{equation}
Therefore, combining the above:
\begin{equation}
\dot \phi - g \eta = - \sigma \partial_x^2 \eta/\rho \qquad \hbox{(at surface)}.
\end{equation}
Inserting the known form of $\phi(x,y,t)$ and $\eta(x,t)$, and factoring out the trigonometric functions,
\begin{equation}
- C \omega \cosh(kd) - ga = \sigma k^2 a /\rho.
\end{equation}
Therefore, using the previously determined formula for $C$, we have
\begin{equation}
+\omega^2 a \coth(kd)/k - ga = \sigma k^2 a /\rho.
\end{equation}
This now leads to the dispersion relation
\begin{equation}
\omega^2  = \left(g + {\sigma \,k^2\over\rho}\right) \; k \; \tanh(kd),
\end{equation}
which we can write in the equivalent form
\begin{equation}
\omega^2  = \left(g\, d+ {\sigma \,d \,k^2\over\rho}\right) \; k^2 \; {\tanh(kd)\over kd}
=  g \, d\left(1+ {\sigma \, \,k^2\over\rho\, g}\right) \; k^2 \; {\tanh(kd)\over kd}.
\end{equation}
Define, in the usual manner,
\begin{equation}
c_0^2 = g \, d; \qquad K_\sigma = \sqrt{\rho g\over\sigma};
\end{equation}
then the dispersion relation reduces to 
\begin{equation}
\omega^2 =  c_0^2 \; k^2 \; \left\{ 1 + {k^2\over K_\sigma^2} \right\} \; {\tanh(kd)\over kd},
\end{equation}
as promised. \hfill $\Box$

\noindent
It is easy to check dimensions:
\begin{equation}
\left[  {\sigma\over\rho c_0^2 d} \right] = { [F]/[L]\over ([E]/[L]^3) [L]} = {[F][L]\over[E]} = 1.
\qquad
\hbox{OK!}
\end{equation}
Similarly one should check appropriate limits:
\begin{itemize}
\item  For $d\to \infty$ one obtains Lamb \S 267 (2) [in the special case $\rho'\to 0$, corresponding to neglecting the density of air as compared to water].
\item  $d\to0$, (more precisely $k\,d\to0$ or $\lambda \gg d$, the shallow water approximation), one obtains
\begin{equation}
\omega^2 = g\,d \,k^2 + {\sigma \,d \,k^4\over\rho} + O(d^3),
\end{equation}
which is a Bogoliubov spectrum, exactly of the form considered above. 
\end{itemize}

\noindent
Returning now to the phase velocity
\begin{equation}
c^2_\mathrm{phase} 
=  c_0^2\;
\left\{ 1 + {k^2\over K_\sigma^2} \right\} \; {\tanh(kd)\over kd}
=  c_0^2\;
\left\{ 1 + \epsilon (k d)^2 \right\} \; {\tanh(kd)\over kd},
\end{equation}
it is easy to see that for low wavenumber
\begin{equation}
c^2_\mathrm{phase} =  c_0^2\;
\left\{ 1 + {3\epsilon-1\over3} \;(k d)^2  - {5\epsilon-2\over 15} \;(k d)^4 + {O}[(k d)^6]   \right\} ,
\end{equation}
implying
\begin{equation}
c_\mathrm{phase} =  c_0\;
\left\{ 1 + {3\epsilon-1\over6} \;(k d)^2  - {45\epsilon^2+30\epsilon-19\over 360} \;(k d)^4 + {O}[(k d)^6]   \right\} .
\end{equation}
The significance of this observation is that by choosing $\epsilon=1/3$, corresponding to $d\; K_\sigma = \sqrt{3}$, one can tune away the lowest-order ``Lorentz violating'' term. For water under normal conditions this corresponds to $d= \sqrt{3}\times (0.27) \hbox{ cm} = 0.47 \hbox{ cm}$. The broader message to take from this discussion is this:  In both analogue models and in quantum gravity phenomenology one should always be careful about the possibility of accidentally tuning away the effect you are looking for.

 The group velocity is relatively messy
 \begin{equation}
{c_\mathrm{group}\over c_\mathrm{phase}} =  
{kd\over\sinh(2kd)} + {1\over2} \left[{1+3\epsilon (kd)^2\over1+\epsilon(kd)^2}\right] \in [1,2],
\end{equation}
though for small wavenumber one has
\begin{equation}
c_\mathrm{phase} =  c_0\;
\left\{ 1 + {3\epsilon-1\over2} \;(k d)^2  - {45\epsilon^2+30\epsilon-19\over 72} \;(k d)^4 + {O}[(k d)^6]   \right\} .
\end{equation}
Comparing with our result for the phase velocity, this is sufficient to guarantee the existence of a low wavenumber hydrodynamic limit.
Note again the significance of $\epsilon=1/3$ in terms of tuning away the lowest ``Lorentz violating'' term. More broadly, $\epsilon<1/3$ corresponds to a dispersion relation that is initially ``subsonic'', while  $\epsilon>1/3$ corresponds to a dispersion relation that is initially ``supersonic''.  (Though as long as $\epsilon>0$ both dispersion relations are eventually ``supersonic'' at sufficiently high wavenumber.)
It is also easy to see that both definitions of the signal velocity agree that $c_\mathrm{signal}=\infty$. 
The messages for quantum gravity phenomenology are twofold:
\begin{itemize}
\item 
The simple division into ``superluminal'' and ``subluminal'' dispersion relations can be seriously misleading, in that even quite physically reasonable dispersion relations can be \emph{both}   ``superluminal'' and ``subluminal''  over different parts of their domain.
\item 
Furthermore, it is not too difficult to obtain physically plausible dispersion relations with freely tunable parameters that can be used to delay if not completely eliminate ``Lorentz violating'' effects. 

\end{itemize}

\section{Discussion}

Many interesting  extensions and modifications of the  general relativity notion of spacetime have 
concrete and well controlled models within the ``analogue spacetime'' framework. (In particular, many interesting dispersion relations can be modelled using water waves.) By and large, this tells us which rocks to start looking under. While there is in no sense any way in which we can currently ``derive'' quantum gravity phenomenology from the ``analogue/ emergent spacetime''  programme, it should nevertheless be emphasized that the ``analogue spacetime'' framework is quite natural and plausible from the point of view of  ``quantum gravity phenomenology''. By looking at specific analogue models we have seen that both ``superluminal'' and ``subluminal'' dispersion relations naturally arise, we have seen that dispersion relations can sometimes be both ``superluminal'' and ``subluminal'' , that the signal velocity can be chosen to be either finite or infinite, that the onset of Lorentz violating effects can sometimes be delayed by fine-tuning,  that dispersion relations can be chosen to ``mimic'' the effects of parity violation,  that ``rainbow metrics'' are ubiquitous, and that pseudo--Finsler metrics are not far behind.  This is a remarkable output for what was originally considered to ``merely'' be an analogy. 

\section*{Acknowledgments}

Various versions of the ideas in this article were also presented at the following conferences:  ``Time and Matter 2007'', August 2007, Lake Bled, Slovenia; at  "Enrageing ideas'', September 2007, Utrecht University, the Netherlands;  and at ``Experimental search for quantum gravity'', November 2007, Perimeter Institute, Canada. MV wishes to thank SISSA/ISAS (Trieste) for ongoing hospitality.


\end{document}